\documentclass[prb,preprint]{revtex4-1} 

\usepackage{amsmath}  
\usepackage{amsfonts} 
\usepackage{graphicx} 
\usepackage{color}

\newcommand{\ain}{\mathrm{A}_\mathrm{in}}
\newcommand{\aout}{\mathrm{A}_\mathrm{out}}
\newcommand{\adp}{\mathrm{ADP}}
\newcommand{\atp}{\mathrm{ATP}}
\newcommand{\phos}{\mathrm{P}_i}
\newcommand{\bin}{\mathrm{B}_\mathrm{in}}
\newcommand{\bout}{\mathrm{B}_\mathrm{out}}
\newcommand{\conc}[1]{[#1]}

\newcommand{\dt}{\Delta t}
\newcommand{\fidl}{F^\mathrm{idl}}
\newcommand{\ftot}{F^\mathrm{tot}}
\newcommand{\kdt}{k_\mathrm{dt}}
\newcommand{\ktd}{k_\mathrm{td}}
\newcommand{\kio}{k_{io}}
\newcommand{\koi}{k_{oi}}
\newcommand{\khatio}{\hat{k}_{io}}
\newcommand{\khatoi}{\hat{k}_{oi}}
\newcommand{\kmn}{k_{mn}}
\newcommand{\knm}{k_{nm}}
\newcommand{\lbra}{\left[}
\newcommand{\lpar}{\left(}

\newcommand{\nain}{N_a^{\mathrm{in}}}
\newcommand{\nainstar}{N_a^{\mathrm{in*}}}
\newcommand{\naout}{N_a^{\mathrm{out}}}
\newcommand{\natot}{N_a^{\mathrm{tot}}}
\newcommand{\nap}{\mathrm{Na}^+}
\newcommand{\nbin}{N_b^{\mathrm{in}}}
\newcommand{\nbout}{N_b^{\mathrm{out}}}
\newcommand{\nbtot}{N_b^{\mathrm{tot}}}
\newcommand{\no}{N_0}
\newcommand{\rbf}{\mathbf{r}}
\newcommand{\rbra}{\right]}
\newcommand{\rpar}{\right)}
\newcommand{\vbf}{\mathbf{v}}
\newcommand{\vin}{V_{\mathrm{in}}}
\newcommand{\vout}{V_{\mathrm{out}}}
\newcommand{\zidl}{Z^\mathrm{idl}}
\newcommand{\ztot}{Z^\mathrm{tot}}
\newcommand{\ztwo}{Z_2}
\begin{document}


\title{Key biology you should have learned in physics class:\\ Using ideal-gas mixtures to understand biomolecular machines}

\author{Daniel M. Zuckerman}
\email{zuckermd@ohsu.edu} 
\affiliation{Department of Biomedical Engineering, Oregon Health \& Science University, Portland, OR 97239}


\date{\today}

\begin{abstract}
The biological cell exhibits a fantastic range of behaviors, but ultimately these are governed by a handful of physical and chemical principles.  Here we explore simple theory, known for decades and based on the simple thermodynamics of mixtures of ideal gases, which illuminates several key functions performed within the cell.  Our focus is the free-energy-driven import and export of molecules, such as nutrients and other vital compounds, via transporter proteins.  Complementary to a thermodynamic picture is a description of transporters via ``mass-action'' chemical kinetics, which lends further insights into biological machinery and free energy use.  Both thermodynamic and kinetic descriptions can shed light on the fundamental non-equilibrium aspects of transport.  On the whole, our biochemical-physics discussion will remain  agnostic to chemical details, but we will see how such details ultimately enter a physical description through the example of the cellular fuel ATP.
\end{abstract}

\maketitle 

\section{Introduction} 

How many of us in the physics community have felt grateful we did not need to master the uncountable lists of special cases which seem to be the sum-total content of biology textbooks?
We physicists prefer a ``beautiful science'' based on just a few principles, from which everything else can be derived with a bit of mathematics.  
%
But is it possible there are just a few basic principles of biology that physicists can learn and use as a springboard to gain non-trivial insight into the field? \cite{nelson2004biological}
Here, I'm not referring to the ultimate principle of evolution, but rather to well-understood physical principles, embodied in equations, that underpin so much of the function of biological cells.  
The idea is not original but builds on prior work, especially that of physicist Terrell Hill \cite{hill2005free} and numerous others. \cite{giancoli1971course,hopfield1974kinetic,schrodinger1944life,phillips2012physical,hoffmann2012life,wolgemuth2011trend,chowdhury2009physicsMachines,chowdhury2013stochastic,nelson2004biological,dill2012molecular,dreyfus2015teachThermo}.


This article will study protein ``machines'' and how they facilitate free energy-driven transport of molecules in and out of a cell.
In essence, every protein performs a job ranging from catalysis (enzymes) to responding to environmental signals (receptors) to locomotion (motors). \cite{alberts2007molecular,chowdhury2013stochastic}
We will focus on \emph{transporters,} which are membrane proteins that transport numerous molecular substrates in or out of the cell, typically in conditions requiring free energy -- i.e., in the direction \emph{opposite} to which the substrate would flow spontaneously if a channel were available.
We will address a number of questions:
What are suitable elementary thermodynamic and kinetic descriptions of transport?
What are the assumptions for these descriptions, and how are they justified?
How can we understand transport in the context of both equilibrium and non-equilibrium physics?
How do coarser descriptions of a system arise in principle from more fundamental microscopic theory?
And finally, in a cell biology context, how far beyond transporters can we expect the simple approaches to be fruitful?

In exploring transporters, we will largely leave aside details of the chemistry and structural biology, but not the physical essentials.
After all, proteins don't ``know'' biology. 
The individual transporter is inanimate, merely a large molecule that obeys the laws of physics and chemistry. \cite{zuckerman2010statistical}
Consistent with its physico-chemical nature, a protein is only capable of four actions: (i) \emph{binding} to another molecule; (ii) \emph{catalysis} of chemical reactions; (iii) \emph{conformational change} of its shape; and (iv) \emph{diffusion} or thermally driven passive motion.
Each of these, especially (i) - (iii), have evolved to operate in chemically precise ways -- e.g., binding to only a small set of molecules or catalyzing a specific set of chemical reactions.  Note that conformational changes typically alter the function of a protein -- e.g., opening or closing a binding cavity or permeation channel.

\begin{figure}[hb]
    \centering
    \includegraphics[width=10cm]{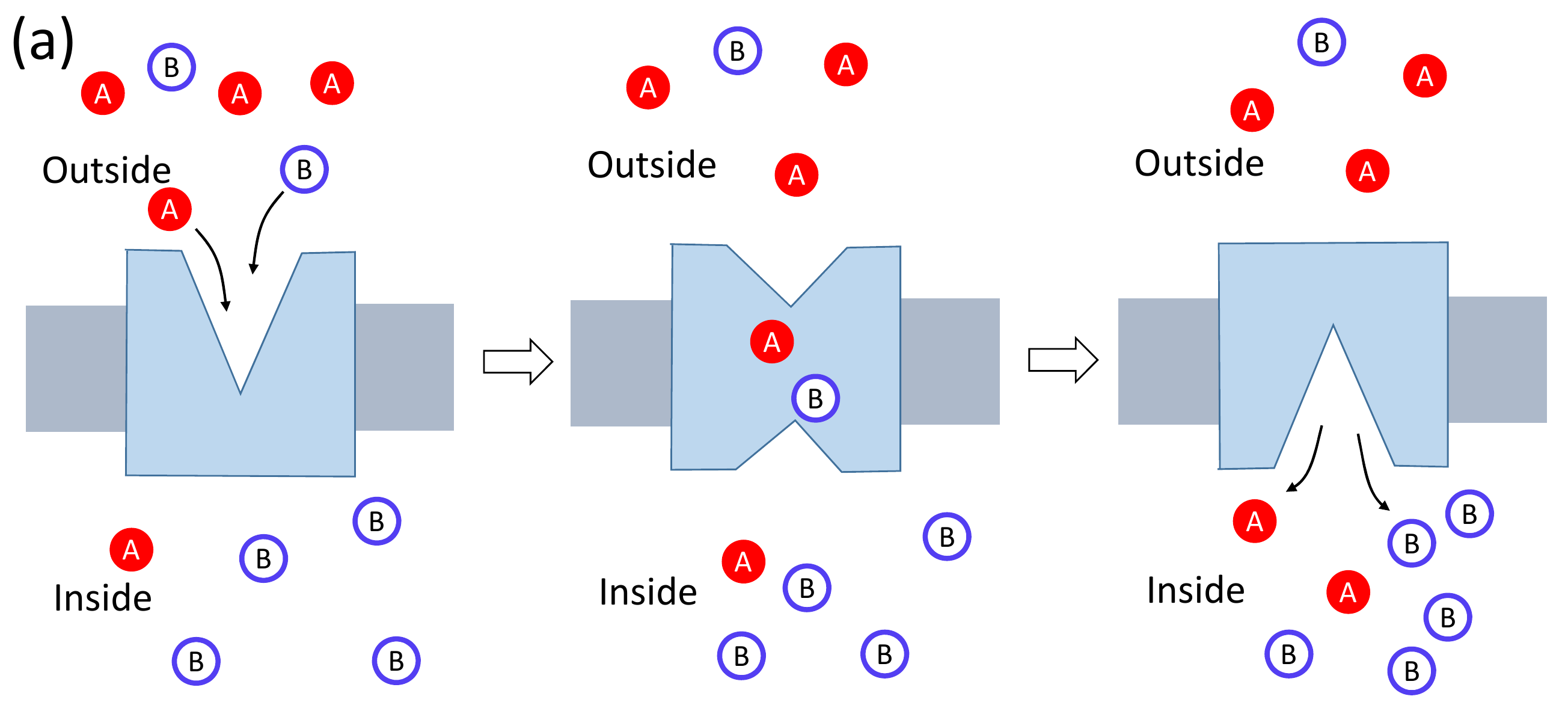}
    \hspace{1cm}
    \includegraphics[width=5cm]{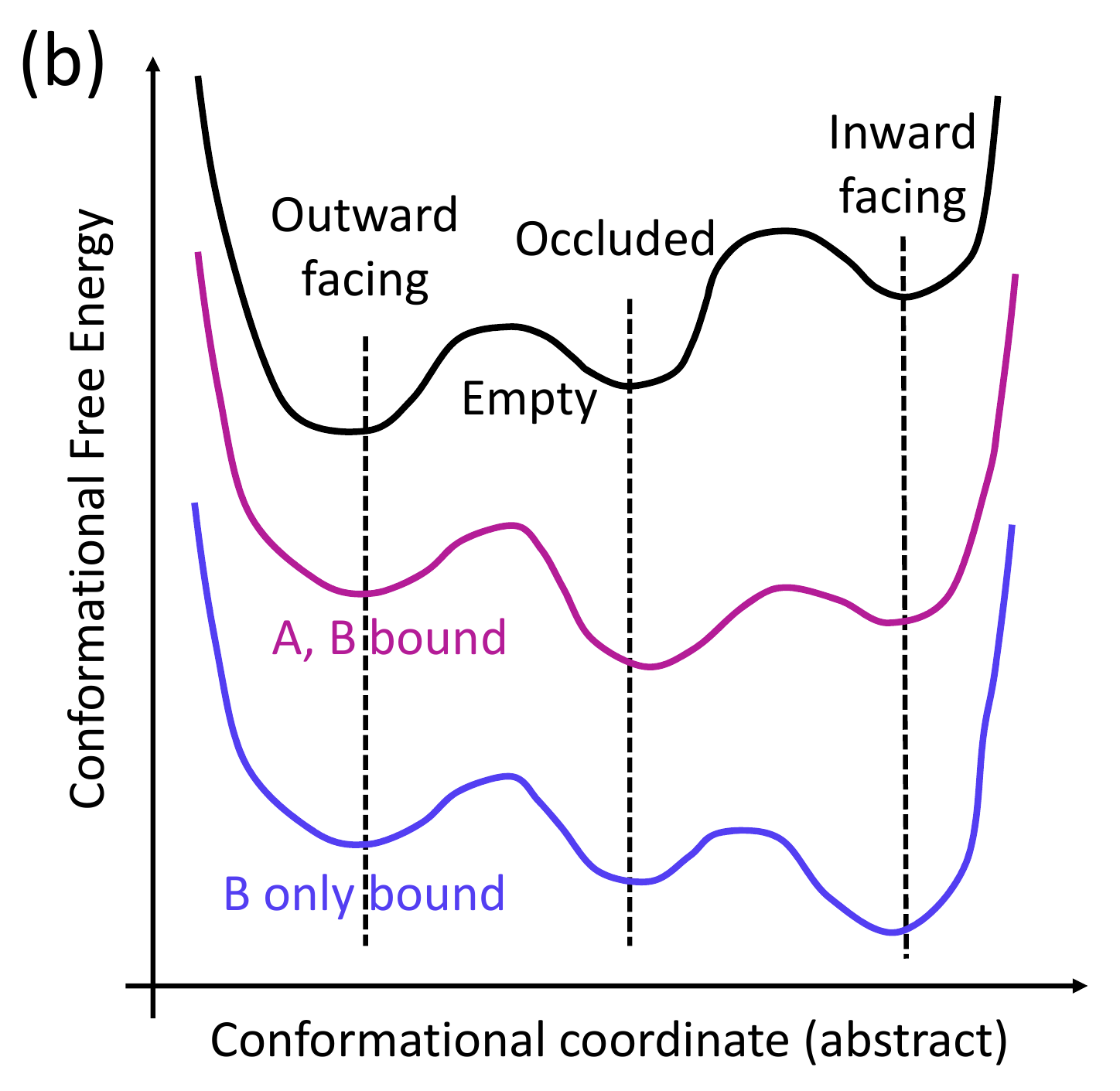}
    \caption{Highly schematic representation of the co-transport/symport process.
    (a) Binding of A and B molecules to the outward-facing conformation of a transporter (light blue) embedded in a membrane (gray) triggers a conformational change that leads to an outward facing conformation where A and B unbind.  The system then resets.  All steps are reversible, with directionality depending on relative concentrations inside and out.  (b) Binding events alter the free energy landscape of the transporter, favoring different conformations during the cycle.}
    \label{fig:symport}
\end{figure}

The real ``magic'' and power of protein machine-like behavior comes from the evolved \emph{coupling} between two or more of the basic actions. \cite{hill2005free}
In the class of transporter proteins to be analyzed here, conformational changes are triggered by binding events.
As sketched in Fig.\ \ref{fig:symport}, if a protein has more than one conformational state, then binding to another molecule -- generically called a \emph{ligand} -- can shift the free energy landscape to favor a different conformation.
Many different chemical mechanisms can cause such a shift, but a simple example would be if the ligand effectively glues together two previously floppy regions of the proteins, perhaps via favorable electrostatics.

The coupling of binding and conformation changes enables transporters to function as molecular machines which use an external source of (free) energy to do work.
The source of free energy often is a gradient of an ion --  more precisely, a difference in chemical potential of the ion across a membrane -- or from an ``activated'' molecule like ATP \cite{alberts2007molecular}.
(As we will see in the Discussion below, even the activation of ATP can only be understood in a thermodynamic context.)
The work done by a transporter typically is to pump a \emph{substrate} molecule to a cellular compartment which represents ``uphill'' motion against its own chemical-potential gradient.

Although an exact theoretical description of transporter function would be highly complex, we can understand the essential ideas using very basic thermodynamics or kinetics.
This is where the ideal gas comes in --- following a long history in the field of biochemistry. \cite{hill2005free,Berg-2002,alberts2007molecular}
We can describe the thermodynamics of transporter systems using the simple equations for a mixture of ideal gases in which there is no potential-energy cost to switch among the components.
The transporter not only couples, say, the ion ``gases'' across the membrane but also can enforce stoichiometric exchange with the gas of the substrate molecule being transported.
That is, if we let A (subscripted as $a$) represent the ion and B (subscript $b$) the substrate, the transporter can be viewed as catalyzing the following exchange across a membrane separating two arbitrary regions which we'll name ``in'' and ``out'' for concreteness:
\begin{equation}
    \aout + \bout \rightleftharpoons \ain + \bin \; .
    \label{chemrxn_sym}
\end{equation}
A process like this where both molecules move in the same direction, as sketched in Fig.\ \ref{fig:symport}, is called ``co-transport'' or ``symport'' and is exemplified by $\nap$ ions (A), which have much higher extra-cellular concentration, being used to drive the import of glucose (B) or another sugar into the cell \cite{Berg-2002}.
Well-studied transporters using this mechanism include Mhp1 which ``salvages'' metabolic precursors from the environment for a microbacterium \cite{weyand2008structure} and vSGLT which imports sugars in a flagellum-powered seawater bacterium \cite{faham2008crystal}.
Numerous other transporter processes occur, as discussed below, but we'll focus solely on \eqref{chemrxn_sym} for clarity.

Our theoretical approach may seem cumbersome at first but will turn out to be fully tractable using undergraduate-level tools.
To start, we consider the Helmholtz free energy $F$ of an ideal mixture of all four components noted in \eqref{chemrxn_sym}:
\begin{equation}
    \ftot(\nain,\naout,\nbin,\nbout; \vin, \vout, T)
    \label{ftotintro}
\end{equation}
where $N$ represents the number of particles of the indicated species and $V$ gives the volume of the subscripted compartment.  
Note that the volumes and temperatures are held constant throughout; only the $N$ values can vary.
However, because of constraints implicit in our model, in fact \emph{there is only one degree of freedom, not four, in Eq.\ \eqref{ftotintro}}!
First, we will assume there are a fixed total number of each chemical species A and B.
Second, the process \eqref{chemrxn_sym} implies that number of inside and outside molecules change in a coupled way.
In the end, as we will see below, the math is greatly simplified.

In addition to the thermodynamic description, it is very valuable to consider the complementary time-dependent viewpoint of chemical kinetics implied by the ``reaction'' \eqref{chemrxn_sym}.
After all, in nature, it is dynamical, microscopic processes which lead to macroscopic thermodynamics and not the other way around. \cite{zuckerman2010statistical}
The most common description in chemistry and biochemistry, known as ``mass action'' kinetics and originated in the 1800s \cite{cornish2012fundamentals}, provides a precise dynamical analogue to the ideal gas because molecular reactants are considered to be non-interacting except for their possibility to transform into the likewise non-interacting products.
This ideality is, of course, an approximation but such a useful one that it is essentially unquestioned throughout biochemistry \cite{Berg-2002}.
The explicitly non-equilibrium nature of a simple kinetic description will echo insights gained from a free-energy picture.

In the context of the physics education literature, the present contribution rests on several points.
First, the author is unaware of a pedagogical, physics-based treatment of biochemical transport in the literature.
Second, the dual thermodynamic/kinetic perspectives offer a model for analyzing other (bio)physical processes.
Third, the material is presented in a manner consistent with mainstream biochemistry, including notation as appropriate, providing the reader a toe-hold into the primary literature of that field.

The remainder of this paper will start by describing biological transporters in sufficient detail to motivate the statistical physics description that follows.
The complementary kinetic description will be given and shown to be consistent with the thermodynamics.
Finally, the discussion section will provide important connections to other biological processes.
Appendices delve into more advanced aspects of the formulation and suggest further undergraduate-level calculations of interest.
The principles described here in the context of transport are quite general and lend insight into numerous biological processes \cite{zuckermanLens}.

\section{Transporters in cell biology}

To set transporters somewhat more broadly in the context of cell biology, not only are there transporters to import every kind of nutrient into the cell but there are numerous ion-only transporters geared toward maintaining the transmembrane electrostatic potential and the well-regulated balance among different ion species in the cell.\cite{alberts2007molecular}
Transporters are also critical in inter-nerve cell communication, not only mediating electrostatic ``action potentials'' but also vital to absorbing neurotransmitters from synapses.\cite{alberts2007molecular}
In short, transporters are not a detail of cell biology, but fundamental.  They are the object of extensive current research.

We focus here on ``secondary active transporters,'' which means the free energy driving the transport is derived from the ``gradient'' of an ion across a membrane.
As noted above, these transporters can use the difference in chemical potentials of the ion inside and outside the cell (or cellular compartment) to power the transport of a specific substrate molecule against its gradient --- from low to high chemical potential.
The full chemical potential includes \emph{all} chemical and physical factors, including electrostatics, but it will not prove necessary to delve into these details here.
Chemical specificity for a given ion and substrate are provided by the particular amino acids in the transporter proteins which tend to provide a fully complementary shape and electrostatic environment. \cite{Berg-2002}.
Roughly speaking, there is a different transporter protein or protein complex for each substrate molecule, although there are notable exceptions. \cite{alberts2007molecular}.
In a thermal environment, unsurprisingly, neither the ligand specificity nor the coupling efficiency of a transporter is perfect,\cite{henderson2019coupling} but we shall put off considering those aspects until the Discussion section.

\section{BASIC THERMODYNAMICS OF TRANSPORT}
\label{sec:theory}

To elucidate the principles of transporter function, we will pursue nearly exact treatments of simple models - a mixture of ideal gases and subsequently mass-action kinetics.
The essence of the ideal-gas theory is long-established in the physics community and broadly accepted (if only implicitly) in the field of biochemistry.
Mass-action kinetics are the cornerstone of quantitative biochemistry \cite{Berg-2002,cornish2012fundamentals} and represent an intuitive physical approach, as will be seen.
The author hopes the typical physicist or biochemist will encounter something new, and hopefully informative, in the combination of approaches to be employed here.

Why are we justified in using ideal-gas theory?  As is often the case when digging into a real-world problem, we start with the simplest sufficient model.  You will see that the ideal gas provides just that.  But less obviously, there are good theoretical reasons to expect some insensitivity to details of molecular interactions.  These points are elaborated upon in Appendix \ref{app:ideal} and are of great importance to readers who want to go a bit deeper into the biophysics.

To tackle transporters, we need to treat two types of molecules A and B, each of which can be inside or outside the cell.
As suggested by the free energy \eqref{ftotintro}, in the ideal-gas picture, we must account for four independent ``gases.''
But of course, these components are not truly independent because transit of an A particle from outside to inside simultaneously changes the A counts in both compartments, not to mention the coupling to B transit via \eqref{chemrxn_sym}.
This coupling can be fully accounted for in a simple additive formulation --- which we shall derive from the full-system partition function later on, for completeness.

\subsection{Refresher: Defining equilibrium, configurations, and states}
Equilibrium is a key reference point in understanding any aspect of statistical physics or biochemistry, even if our main goal is to understand non-equilibrium phenomena.
As we'll see below, equilibrium theory will enable us to understand almost all of the free-energetics of our system.

To discuss equilibrium, we first need some nomenclature.
A \emph{configuration} is defined to be a single point in phase space -- i.e., the set of all particle positions and momenta for a classical system \cite{zuckerman2010statistical}.
On the other hand, a \emph{state} is taken to be a collection of configurations, which can be defined in almost any way that is convenient although not all definitions are equally useful.\cite{chodera2007automatic}
Somewhat confusingly, a ``steady state'' does not refer to a chosen set of configurations but rather to a constant-in-time distribution of \emph{all} configurations.

Equilibrium is a special steady state, defined by more than time-invariant properties. \cite{hill2005free,zuckerman2010statistical}
Not only must the population of each species (e.g., B outside the cell) and its spatial distribution be constant in time, but further there must be no flows in the system, on average.
That is, if we observe the system over a long period of time, although a given species may get transported numerous times, there should be an equal number of transits in each direction.
Likewise, although molecules may diffuse spatially, there should be no ``rivers'' of uni-directional motion.
All this can be encapsulated in the notion of \emph{detailed balance}, which means an equal and opposite average flow of particles/probability per unit time between any pair of configurations.
If you think about it, detailed balance implies unchanging probabilities because each configuration gains as much as it loses.
Note that if detailed balance holds among ``microscopic'' configurations, then it also must hold among coarser states. \cite{zuckerman2010statistical}

In equilibrium, the probability of a system configuration is governed by its Boltzmann factor (see below) a property which is critical to defining the partition function and the free energy.

\subsection{Refresher: A single ideal gas}
To introduce some notation and remind ourselves of ideal-gas statistical mechanics, let's start from the partition function for an equilibrium system of $N$ classical, non-interacting particles of mass $m$ at temperature $T$ confined to a fixed volume $V$.
A partition function is just a sum and/or integral over the equilibrium Boltzmann factors for every possible system configuration defined by positions and velocities:
consult your favorite statistical mechanics book for reference (e.g., Refs.\ \onlinecite{reif2009fundamentals, hill1986introduction, zuckerman2010statistical}).

Fortunately, the ideal gas partition function can be written down and evaluated easily.
Because ideal particles experience no forces from one another or the container walls, the potential energy is a constant (taken to be zero) independent of the positions of the particles.
Hence the total energy needed for the partition function is solely kinetic.
Letting $\rbf_i = (x_i, y_i, z_i)$ denote the position of particle $i$ and analogously defining $\vbf_i$ as the velocity vector with magnitude $v_i$, we can write the classical ideal gas partition function and evaluate the integrals exactly yielding \cite{hill1986introduction, reif2009fundamentals, zuckerman2010statistical}
\begin{align}
    \zidl(N, V, T) &= \frac{1}{N!} \lpar \frac{m}{h} \rpar^{3N} 
    \! \int d\vbf_1 \cdots d\vbf_N
    \int d\rbf_1 \cdots d\rbf_N \,
    \exp \! \lbra-\sum_{i=1}^N (1/2) m \left. v_i^2 \right/ k_B T \rbra \nonumber \\
    &= \frac{1}{N!} \lpar \frac{V}{\lambda^{3}} \rpar^N \; ,
    \label{zidlone}
\end{align}
where $h$ is Planck's constant, $k_B$ is Boltzmann's constant, and $\lambda(T) = h/\sqrt{2\pi m k_BT}$.

The Helmholtz free energy is then obtained as
\begin{align}
    \fidl(N,V,T) &= -k_B T \ln \zidl 
              \simeq k_BT \lbra N \ln N - N - N \ln ( V / \lambda^3) \rbra \nonumber \\
           &  \simeq N k_B T \ln \lpar \frac{N}{V/\lambda^3} \rpar \; ,
           \label{fidlone}
\end{align}
where we have employed Stirling's approximation as usual. \cite{reif2009fundamentals, hill1986introduction, zuckerman2010statistical}
Importantly for biochemical applications, note that the fundamental dependence here is $\fidl/N \sim \ln (N/V)$ --- i.e., the log of the number density $N/V$, also called \emph{concentration}.
We shall be assuming constant temperature throughout, appropriately for biochemistry, so the temperature dependence is not directly pertinent.

\subsection{The simplest mixture: Two ideal gases exchanging across a membrane}
\label{sec:two-gases}

\begin{figure}
    \centering
    \includegraphics[width=6cm]{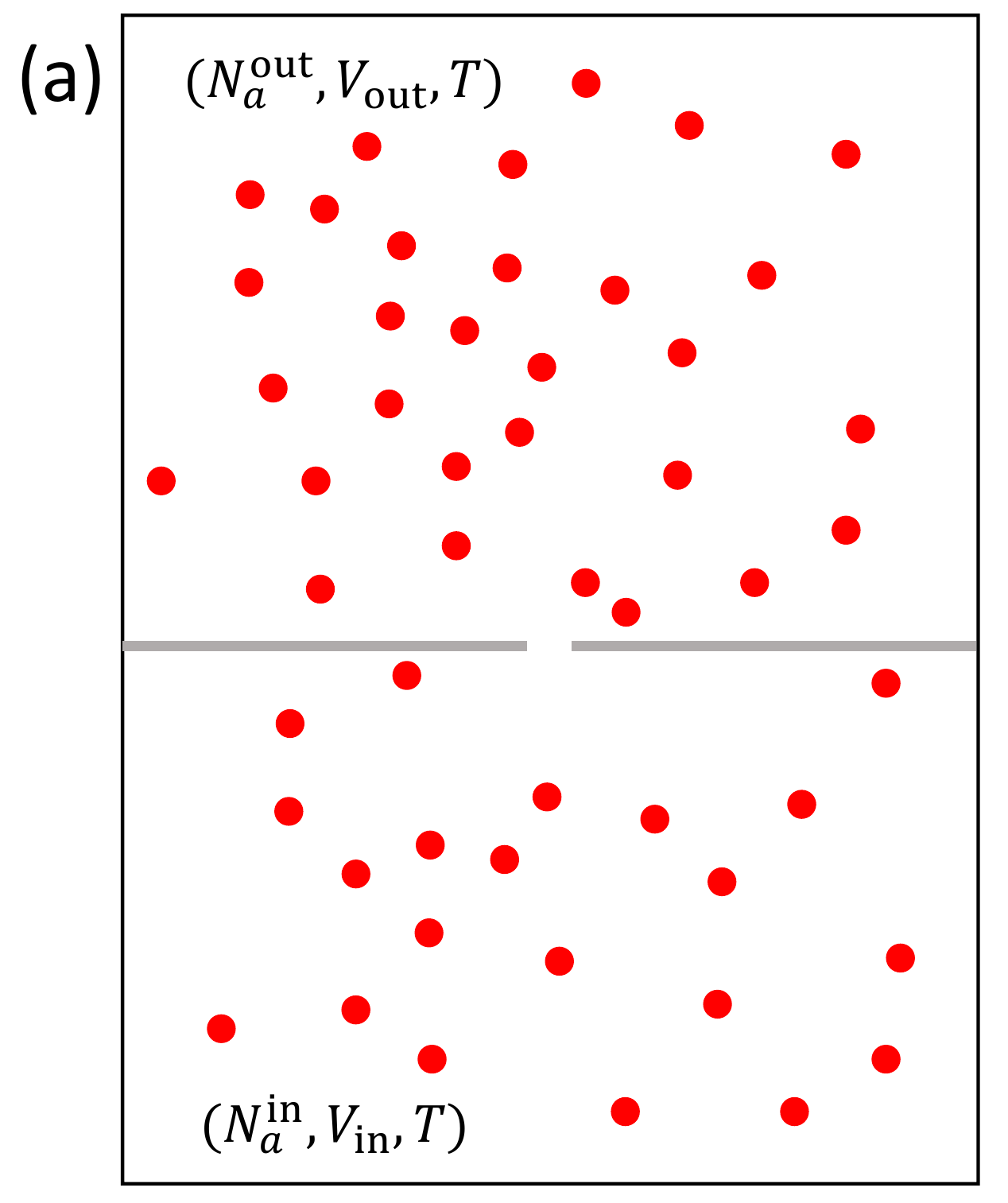}
    \hspace{1cm}
    \includegraphics[width=7cm]{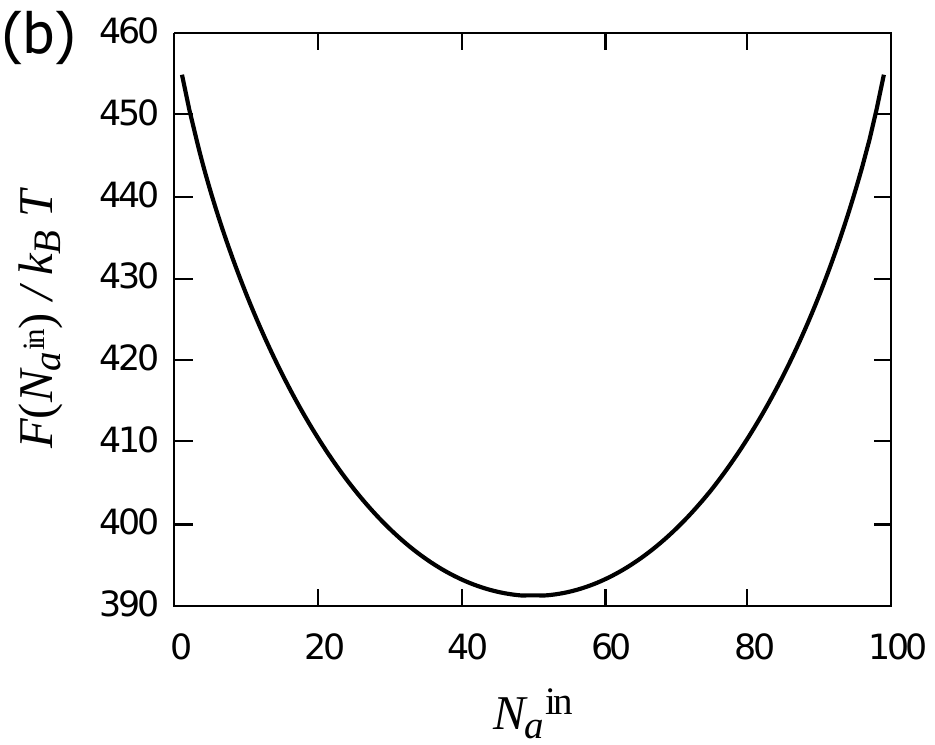}
    \caption{Free energy minimization for a single ideal gas in a container with a permeable divider.  (a) The ideal gas particles are divided between two compartments, separated by a permeable divider which enables the system to sample all possible particle allotments between the compartments.  (b)  The free energy \eqref{fidltwo} is plotted as a function of $\nain$ with $\natot=100$, $\vin=\vout$ and $\vin/\lambda^3 =1$ for convenience.}
    \label{fig:ideal_two}
\end{figure}

As a second step toward modeling a biological transporter, we'll increase the complexity of our system incrementally  by considering two separated ideal gases of the same molecule type (A) which can exchange particles across a membrane, as in Fig.\ \ref{fig:ideal_two}.
To suggest a connection with cell behavior, we'll call the two compartments ``inside'' and ``outside'' with volumes $\vin$ and $\vout$ as well as corresponding particle numbers $\nain$ and $\naout$ constrained to sum to $\natot$.
We can explicitly write out the total free energy as the sum of the two ideal-gas free energies from \eqref{fidlone}; the additivity is justified in Appendix \ref{sec:partition}.  We obtain
\begin{align}
    \ftot_a & = \nain \, k_B T \ln \lpar \frac{\nain}{\vin/\lambda^3} \rpar +
              \naout \, k_B T \ln \lpar \frac{\naout}{\vout/\lambda^3} \rpar \nonumber \\
          & = \nain \, k_B T \ln \lpar \frac{\nain}{\vin/\lambda^3} \rpar +
              (\natot - \nain) \, k_B T \ln \lpar \frac{\natot - \nain}{\vout/\lambda^3} \rpar \label{fidltwo} \\
          & = \ftot_a(\nain; \natot, \vin, \vout, T) \; , \nonumber
\end{align}
where we used $\naout = \natot - \nain$.
Because $\natot, \vin, \vout, T$  are considered constants, this free energy has only a single adjustable ``parameter'' $\nain$ as shown in Fig.\ \ref{fig:ideal_two}.

Our system will self-adjust, via in-out exchange of A particles, until a free energy minimum is reached. \cite{reif2009fundamentals}
It is straightforward to calculate this equilibrium point by setting the derivative of $\ftot_a$ to zero.
We first find that
\begin{align}
    \frac{d \ftot_a}{d \nain} &=
    \, k_B T \lbra \ln \lpar \frac{\nain}{\vin/\lambda^3} \rpar + 1 \rbra 
    - \, k_BT \lbra \ln \lpar \frac{\natot - \nain}{\vout/\lambda^3} \rpar + 1 \rbra \; ,
\end{align}
where we have used the total, not partial, derivative notation here because we have explicitly included \emph{all} dependence on $\nain$.
Setting the derivative to zero then yields
\begin{equation}
    \frac{\nain}{\vin} = \frac{\natot - \nain}{\vout} = \frac{\naout}{\vout} \; 
    \hspace{1cm} \mbox{(at equilibrium)},
    \label{eqconctwo}
\end{equation}
which is \emph{a condition of equal inside and outside concentrations.}
This result should not be surprising, since there is no driving force or interaction favoring inside vs.\ outside.
Nevertheless, our calculation illustrates a generally useful procedure: we can usefully combine ideal-gas free energies, write them in terms of a single parameter, and then minimize the result to find the equilibrium point.
We can repeat these steps in more complicated scenarios.

There is still one very important bit of physics we should take away from this calculation.
In particular, the principle that the free energy is minimized with respect to an adjustable parameter ($\nain$, in our case) immediately tells us that the free energy is higher at (non-equilibrium) $\nain$ values which do not satisfy \eqref{eqconctwo}.
That is, \emph{there is usable energy available when the system is away from the minimum} but none is available at the minimum---i.e., at equilibrium---itself, so long as $\natot, \vin, \vout, T$ are held constant.
See Fig.\ \ref{fig:ideal_two}.
After all, the free energy literally means the energy \emph{available} to do work \cite{reif2009fundamentals, zuckerman2010statistical}, but more precisely it is the available energy referenced to the minimum accessible value.

As we will see below, if different components of a system are suitably coupled, the free energy of one component can drive work done on the other.
Active transport across a membrane is the perfect case in point!
Biological cells have developed myriad ways of ``transducing'' free energy to perform the tasks necessary for life \cite{hill2005free, zuckermanLens}, and the Discussion will sketch a few examples beyond transport.

\subsection{Transport thermodynamics: A four-part mixture of particle-exchanging ideal gases}

We are now ready to construct the ideal thermodynamics for the 1:1 biological transporter executing the process \eqref{chemrxn_sym}.
Although there are four components in the free energy \eqref{ftotintro} to account for both species inside and outside the cell, our calculations turn out to be quite simple.
For mathematical convenience, we will make several simplifications which do not affect the key physics we wish to understand.
We assume that both species, A and B, have the same total number of particles,
\begin{equation}
    \natot = \nain + \naout = \nbtot = \nbin + \nbout \; ,
    \label{abtot}
\end{equation}
and that each species has the same mass, so that $\lambda_a = \lambda_b = \lambda$.
We will enforce the stoichiometric coupling \eqref{chemrxn_sym} of our transporter---that one and only one of each type of particle is transferred at one time, in either direction---by constraining the particle counts inside and outside to change in lockstep 
\begin{equation}
    \nbin = \nain + \no \; ,
    \label{offset}
\end{equation}
where $\no$ is an offset, akin to an initial condition, that ultimately will prove critical to understanding the biological transporter.
The preceding assumptions are very convenient and will not change the fundamental conclusions in any way.

Once again, we write the full free energy \eqref{ftotintro} as the sum of individual ideal-gas expressions \eqref{fidlone}.
Because of the constraints just noted, $\ftot$ will depend only on a single adjustable parameter, which we choose to be $\nbin$ because it will most directly help us understand transport.
(We could have chosen any of the other three particle numbers: there is no effect on the final result.)
Summing the free energies based on Appendix \ref{sec:partition} and substituting based on constraints, we have
\begin{align}
    \ftot 
    &= \nain \, k_B T \ln \lpar \frac{\nain}{\vin/\lambda^3} \rpar
              \naout \, k_B T \ln \lpar \frac{\naout}{\vout/\lambda^3} \rpar \nonumber \\
          & \hspace{1cm}   + \nbin \, k_B T \ln \lpar \frac{\nbin}{\vin/\lambda^3} \rpar +
              \nbout \, k_B T \ln \lpar \frac{\nbout}{\vout/\lambda^3} \rpar \nonumber \\
    &= (\nbin-\no) \, k_B T \ln \lpar \frac{\nbin-\no}{\vin/\lambda^3} \rpar \nonumber \\
           & \hspace{1cm} + (\nbtot-\nbin+\no) \, k_B T \ln \lpar \frac{\nbtot-\nbin+\no}{\vout/\lambda^3} \rpar \nonumber \\
          & \hspace{1cm}   + \nbin \, k_B T \ln \lpar \frac{\nbin}{\vin/\lambda^3} \rpar 
                  + (\nbtot - \nbin) \, k_B T \ln \lpar \frac{\nbtot - \nbin}{\vout/\lambda^3} \rpar \nonumber \\
    & = \ftot(\{N_i\}, \vin, \vout, T; \nbin) \; ,
              \label{ftotnbin}
\end{align}
where $\{N_i\}$ is shorthand for the full set of particle variables.
The adjustable parameter $\nbin$ is set off from others to remind us of its importance.

Following the procedure of the preceding section, we minimize $\ftot$ with respect to the adjustable parameter.
Differentiation yields
\begin{align}
    \frac{ d \ftot }{ d \nbin }
    &= k_B T \bigg[ 
    \ln \lpar \frac{\nbin - \no}{\vin/\lambda^3} \rpar + 1  - \ln \lpar \frac{\nbtot - \nbin + \no}{\vout/\lambda^3} \rpar - 1
    \nonumber \\
    &  \hspace{1cm} +
    \ln \lpar \frac{\nbin}{\vin/\lambda^3} \rpar + 1 - \ln \lpar \frac{\nbtot-\nbin}{\vout/\lambda^3} \rpar - 1  
    \bigg] \; .
\end{align}
We obtain a deceptively simple but key result by setting the derivative to zero, and substituting for more interpretable $\{N_i\}$ variables: 
\begin{equation}
    \frac{ \nain / \vin }{ \naout / \vout } =  \frac{ \nbout / \vout }{ \nbin / \vin} \; .
    \label{eqconcfour}
\end{equation}
Quite simply, this relation implies that if we increase the \emph{outside} concentration of A (our ``driving'' molecule) then the system tends to an equilibrium with higher \emph{inside} concentration of B (the ``substrate'' we hope to see transported).
This is the thermodynamic signature of a pump --- a.k.a. a biochemical transporter!
We have obtained a simple physics result, arguably one that is obvious in retrospect since the coupling \eqref{chemrxn_sym} means that A and B will move together, but it is of profound importance in biochemistry.
Further, as will be described in the Discussion, the analysis presented here is paradigmatic for biochemical processes that are much more challenging to intuit.

We can re-frame the preceding comments on driving in terms of stored free energy, building on our initial discussion in Sec.\ \ref{sec:two-gases}.
The minimum free energy condition \eqref{eqconcfour} represents equilibrium, and if we displace the system away from this minimum the system will tend to move back toward it.
Such driving can be seen as a consequence of the system having a higher free energy, above the minimum.
That excess energy can be used to do work, namely, pumping B molecules from outside to inside --- \emph{a process which will occur even if the inside concentration of B already exceeds the outside concentration of B!}
Pumping will occur, for example, if the left-hand ratio of A concentrations in \eqref{eqconcfour} starts at a value of $1/10$ while the right-hand B ratio starts at $1/2$.
Based on the transporter's 1:1 coupling \eqref{chemrxn_sym}, both A and B molecules will move from outside to inside until the ratios of \eqref{eqconcfour} match.

In the bigger picture, given the simple physics involved, we should avoid the mistaken notion that the biochemistry of transporters is trivial.
In fact, we have taken the non-trivial for granted in our whole development.
The ``magic'' of the biochemistry lies firstly in the function of a protein (or protein complex) which actually enforces the reaction/condition \eqref{chemrxn_sym}.
Second, the cell continually uses its energy resources to maintain a highly non-equilibrium gradient of sodium ions across the plasma membrane, effectively a battery \cite{alberts2007molecular}.
Once these \emph{highly} non-trivial features are arranged, it is fair to say the rest is simple.

On a more technical level, note that the equilibrium result \eqref{eqconcfour} is not generally the same as the prior equilibrium finding \eqref{eqconctwo} applied to both A and B species separately.
Because the transporter enforces stoichiometric movement of particles, \eqref{eqconcfour} is a \emph{constrained equilibrium point}, rather than a global equilibrium, as explained further in Appendix \ \ref{sec:constrained}.

\section{Kinetic description of transport}
\label{sec:kinetics}
We can gain a deeper understanding of transport, which fundamentally is a non-equilibrium phenomenon, using a chemical-kinetics description.
This will not involve any chemical details or structures of biomolecules, but rather the simplest possible time-dependent description of transport via basic differential equations.
The approach we take is completely standard \cite{cornish2012fundamentals, beard2008chemical, hill2005free} and is sometimes called a ``mass-action'' description which refers to the simple concentration dependencies assumed for transition probabilities.

Mass-action kinetics, as we will see, are a fairly precise analog of ideal-gas thermodynamics in the sense that both assume particles are non-interacting and both lead to the same equilibrium point.
However, the kinetic picture assumes a ``reaction'' (transport, in our case) probability per unit time that depends on the product of concentrations of any ``reactants''.
In other words, the particles don't interact ... until they do.
Further confirmation for the ideal-gas/mass-action relationship comes from analyzing our kinetic description at its stationary point, which yields the same relationship for the equilibrium concentration as was derived from the free energy picture.

The starting point for a kinetic description will be the ``reaction'' \eqref{chemrxn_sym} performed by our transporter, which we repeat here for convenience:
\begin{equation}
\aout + \bout \rightleftharpoons \ain + \bin \; .
\tag{\ref{chemrxn_sym}}   
\end{equation}
The mass-action formulation quantifies the time dependence of concentrations (number densities) of the chemical species, which will be characterized via biochemical notation,
\begin{equation}
    \conc{X} = N_X / V \; ,
\end{equation}
in \emph{molar} (M) units.
We also require the forward and reverse reaction rate constants for \eqref{chemrxn_sym}, $\kio$ and $\koi$, which are reaction probabilities per mole per unit time \cite{cornish2012fundamentals, zuckerman2010statistical} --- and have units of (M s)$^{-1}$.
Rate constants are independent of both time and concentration, by assumption.

We are almost ready to write down the key equation.
First note that in the mass-action picture, the \emph{overall rates} for the two directions of the ``reaction'' \eqref{chemrxn_sym} are
\begin{equation}
    \conc{\aout} \conc{\bout} \, \koi \hspace{0.25cm} \mbox{(out$\to$in)}
    \hspace{1.5cm}
    \conc{\ain} \conc{\bin} \, \kio \hspace{0.25cm} \mbox{(in$\to$out)} \; ,
    \label{rates}
\end{equation}
where you should note the distinction between overall rate and rate constant.
These expressions can be understood intuitively. \cite{phillips2012physical, zuckerman2010statistical}.
Consider first a single A molecule in a fixed volume $V$, along with $N_b$ fully independent B molecules.
The probability of an A-B encounter is proportional to to $N_b/V = \conc{\mathrm{B}}$.
If we now include a total of $N_a$ A molecules, that encounter probability will increase by a factor of $N_a/V = \conc{\mathrm{A}}$, at least in the limit of small $N_a$.
The mass-action assumption is that the overall reaction probabilities (per mole) are of the simple product form \eqref{rates} \emph{regardless of the A and B concentrations.}
Thus, the reaction probability can always increase regardless of whether, for instance, B molecules are already fully surrounded/caged by A molecules in a spatially realistic picture --- an assumption which essentially mirrors the non-interacting nature of ideal particles which can occupy the same location without energetic cost.
Also implicit in the mass-action expressions \eqref{rates} is an isotropic assumption that the concentrations remain uniform in space -- i.e., that any spatial fluctuations rapidly dissipate via diffusion.

The governing mass-action kinetic equation for a given species then reflects the difference between the overall forward and reverse rates \eqref{rates}.
We will focus on the ``substrate'' $\bin$ because biochemical pumping should increase this concentration under cellular conditions:
\begin{equation}
    \frac{d \conc{\bin} }{dt}
    = \conc{\aout} \conc{\bout} \, \koi
    - \conc{\ain} \conc{\bin} \, \kio \; .
    \label{dbdt}
\end{equation}
The first term on the left is the rate of ``formation'' of $\bin$ in the mass-action formulation, while the second term is the rate of removal/destruction.
Eq.\ \eqref{dbdt} is the \emph{only} differential equation needed for our system because the analogous equations for the other three species can readily be derived from \eqref{dbdt} based on the relationships \eqref{abtot} and \eqref{offset} between the species counts.
For example, $d\conc{\ain}/dt = d\conc{\bin}/dt$ and $d\conc{\bout}/dt = -d\conc{\bin}/dt$.

Note that a chemical-kinetics equation such as \eqref{dbdt} is a \emph{deterministic and averaged} description, which is sufficient for many purposes such as ours.  
However, the actual behavior will be stochastic, and a given system is not expected to precisely follow the average.
Such fluctuating outcomes will instead be governed by a chemical master equation, as discussed in Appendix \ref{app:master}.

We turn to the stationary (steady-state) behavior of \eqref{dbdt}, which will turn out to constrain the rate constants.
Setting the time derivative to zero and re-arranging terms leads to a relation among the steady-state concentrations:
\begin{equation}
    \frac{ \conc{\ain} }{ \conc{\aout} } = \frac{ \conc{\bout} }{ \conc{\bin} } \, \frac{ \koi }{ \kio }
    \hspace{1cm} \mbox{(at steady state = equilibrium)} \; .
    \label{steadyconc}
\end{equation}
Although a steady-state does not necessarily correspond to equilibrium, in our case it does because there are no inputs or outputs of energy or matter to our system \cite{hill2005free, zuckermanLens}.

It turns out we implicitly have more information about the ratio of rate constants occurring in \eqref{steadyconc}.
We have assumed our A and B particles are non-interacting and further do not experience any external field (e.g., electrostatic) which might discriminate inside vs.\ outside.
(Had there been such a field, there would have been an energy term for it in our free energy formulation.)
Therefore the equilibrium point cannot favor inside or outside and we must $\kio = \koi$.
In a kinetic picture, this means the transporter does not favor one direction over the other, consistent with out ideal gas perspective.

Once we recognize that $\koi/\kio = 1$ by our prior assumptions, we see that \eqref{steadyconc} is equivalent to our previous result \eqref{eqconcfour} derived thermodynamically.
This helps to confirm the hypothesized relationship between mass-action kinetics and ideal-gas thermodynamics.

\begin{figure}
    \centering
    \includegraphics[width=10cm]{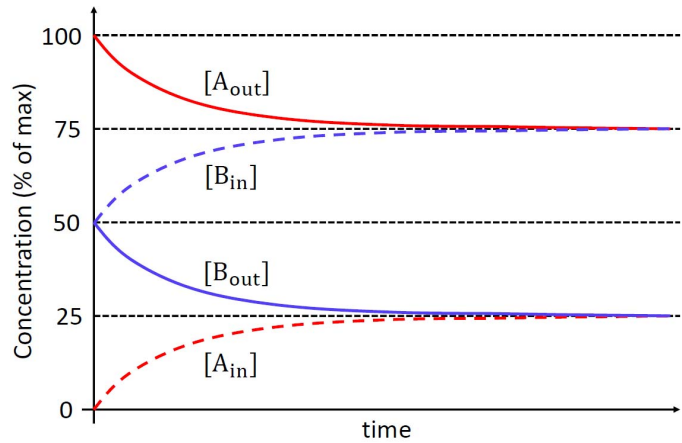}
    \caption{Relaxation to equilibrium in the symporter system (schematic).  Initially, all A molecules are outside, while B molecules are split evenly between inside and outside.  The driving force from the A molecules, which are further from their own equilibrium, pumps B molecules from outside to inside until the constrained equilibrium condition \eqref{eqconcfour} is satisfied.  Here, we assume $\natot = \nbtot$ and $\vin = \vout$.}
    \label{fig:conc_vs_time}
\end{figure}

Although our differential equation \eqref{dbdt} naturally allows examination of transient behavior, that will not be our focus here.
We'll simply point out that if the system is initiated away from its steady state, it will relax toward that steady state over time.
See Fig.\ \ref{fig:conc_vs_time}.
The relaxation will be exponential in a simple system like ours.

In the context of biochemical transport, which may reasonably be considered to occur at steady state, \cite{zuckermanLens} it is very instructive to consider \emph{non}-equilibrium steady states driven by processes external to our system.
In particular, for transporters, the driving A molecule (often an ion) is generally maintained far from the equilibrium point it would attain uncoupled to B because it is continually pumped out of the cell using free energy, described below, from ATP hydrolysis \cite{alberts2007molecular, zuckermanLens}.
Most precisely, we can say that in a cellular context, the chemical potential of A is much higher outside than inside, so there is a thermodynamic driving ``force'' \cite{hill2005free} on A in the out$\to$in direction.
In general, the chemical potential depends on everything in a molecule's environment, including electrostatics and van der Waals interactions. \cite{hill1986introduction}
In our ideal system with no interactions, however, only the species concentrations affect the chemical potential, so we model a driving force by assuming the outside concentration of A greatly exceeds the inside value: $\conc{\aout} \gg \conc{\ain}$.

What happens to B when there is a driving force on A?
The answer is intuitive: because B transport is coupled to A via \eqref{chemrxn_sym}, then B will also be driven from outside to inside the cell.
The key point is that this can occur \emph{even when B is driven against its own gradient} --- from lower to higher chemical potential.
This driving is readily quantified by returning to the fundamental differential equation, armed with the knowledge that $\kio = \koi$.
Based on \eqref{dbdt}, $\conc{\bin}$ will increase whenever right-hand side is positive:
\begin{equation}
    \frac{ \conc{\aout} }{ \conc{\ain} } > \frac{ \conc{\bin} }{ \conc{\bout} } \; .
\end{equation}
Thus, if molecule A is sufficiently far from its own equilibrium of equal concentrations \eqref{eqconctwo}, it can drive B from low to high concentrations.
This is the essence of gradient-driven transport, and is easily appreciated simply based on the sign of the time-derivative for the species of interest.

\section{Discussion}
\label{sec:disc}

\subsection{Yes, physics matters}
The primary goal of this article, broadly speaking, is to introduce a physics-trained audience to essential cell biology concepts framed strictly using undergraduate-level physics.
The take-home message should be that physics is essential to understanding cell biology, a point that has long been appreciated at least implicitly by subsets of the biological community --- e.g., the fields of biochemistry \cite{Berg-2002}, bioenergetics \cite{nicholls2013bioenergetics}, and some cell biology authors \cite{alberts2007molecular}.
Advanced physics is not required to understand some of the most important phenomena, \cite{nelson2004biological} and further examples are given below.
The humble ideal gas has great power in the right context.

At the same time, some topics which are under-emphasized in typical undergraduate, and even graduate, physics curricula have been featured.
These include: (i) the value of reciprocal kinetic and thermodynamic descriptions; (ii) the fundamental importance of non-equilibrium (NE) phenomena and the ease with which NE basics can be presented; and (iii) insight into the meaning and approximation of that taken-for-granted phrase, ``free energy minimization.''
In other words, the application of familiar ideas to a new problem can deepen our understanding of old material.

By no means is this article intended to be a survey or overview of the importance of physics in understanding biology, nor a presentation of the most interesting biology one can understand with physics.
Far from it.
The hope was to go deep enough into a single problem for readers to appreciate that there is a deep and substantial role for physics in biological study.
However, it's worth considering which additional problems can be addressed with the simple ideas discussed here.


\subsection{Beyond simple co-transport}
We have focused our attention on a 1:1 symporter, or co-transporter, which carries out the process \eqref{chemrxn_sym}, but the cell uses many variations on this theme.
Other transporters fall into the class of ``antiporter'' or exchanger, which generate a contrasting process:
\begin{equation}
    \aout + \bin \rightleftharpoons \ain + \bout \; .
    \label{chemrxn_anti}
\end{equation}
The treatment of 1:1 antiport is analogous to our analysis above. \cite{zuckermanLens}

In both symporters and antiporters, different stoichiometries occur, so that two ions (A) might be required to transport one sugar molecule (B), for instance. \cite{alberts2007molecular}
All such transporters, which employ free energy stored in the inside-outside chemical potential difference, are called ``secondary active transporters'' to distinguish them from ``primary'' transporters that hydrolyze ATP to perform transport.
Primary active transporters may also involve multiple substrates in different stoichiometries \cite{Berg-2002, alberts2007molecular}.

Beyond stoichiometric variation, there is a growing awareness that transporters may not always function in simple stoichiometric fashion or by simple mechanisms. \cite{henderson2019coupling, bazzone2017loose, robinson2017new}
That is, the ratio of substrate to ion (B to A) molecules moved per transport cycle may not need to be an integer.
Mechanistically, this likely results from the ``slippage'' phenomenon quantified by physicist Terrell Hill in his seminal book. \cite{hill2005free}
In other words, in a detailed map of the network of possible processes occurring within a transporter, some may result in apparently futile leakage of either substrate or ion down its electro-chemical gradient.
Slippage, as well as the question of mechanistic heterogeneity, \cite{robinson2017new} are active research topics.


It is fair to say there are processes far more remarkable than transport occurring in the cell which can be modeled using a straightforward physical approach.
Perhaps the most exciting is a phenomenon called ``kinetic proofreading'' (KP) which was independently discovered by a physicist and a biochemist \cite{hopfield1974kinetic, ninio1975kinetic}.
It is fair to say that KP is one of the fundamental ``secrets of life'' \cite{chen2018stochastic}, but unfortunately remains too much of a secret: it is not a textbook subject, and is little known in either the biological or physical communities.

Quite simply, KP can be described as a generic strategy of using free energy to preserve information, or more precisely, to achieve higher biochemical discrimination than would be possible without the extra energy use.
For example, KP is what permits our cells to translate proteins from mRNA with an error rate of about $10^{-4}$ instead of $10^{-2}$.
Without it, you would not be reading this article; our species could not exist.
KP can be understood using undergraduate-level physics akin to what is described above \cite{zuckermanLens, chen2018stochastic}, and it also has received more general physics treatments. \cite{murugan2014discriminatory}
This is a great topic for anyone seeking to delve deeper into physical biology.

\subsection{Last word: Chemical details and the example of ATP free energy}

\begin{figure}
    \centering
    \includegraphics[width=8cm]{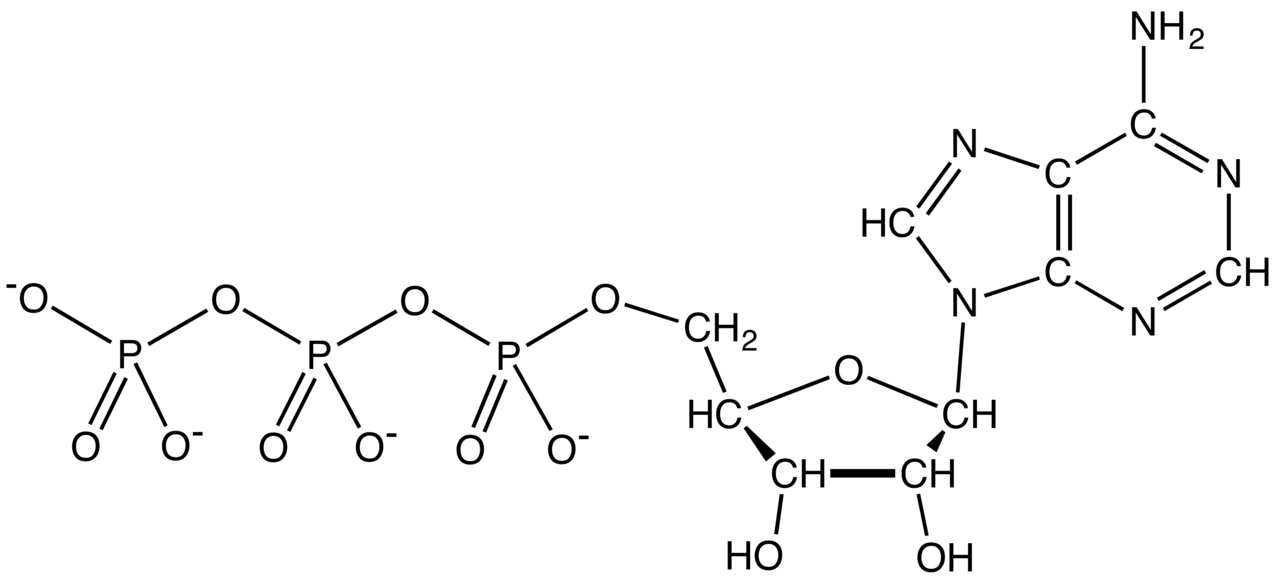}
    \caption{The universal fuel of the cell, ATP.  The bonds connecting the charged phosphate groups are said to be high-energy.  However, the true source of free energy obtained from ATP is due to the concentrations of ATP and its hydrolysis products being maintained far from equilibrium.}
    \label{fig:atp}
\end{figure}

Adenosine tri-phosphate (ATP, Fig.\ \ref{fig:atp}) is surely one of the most important and most misunderstood molecules.
It plays a key role in transport, as the driver of a wide class of ``primary'' active transporters (which are \emph{not} driven by ion gradients). \cite{alberts2007molecular}
We learn in high school that ATP is the ``fuel'' of the cell, which is roughly true but somewhat misleading.
Our earlier discussion was somewhat more precise in referring  to ATP as ``activated'' \cite{alberts2007molecular, zuckermanLens}, but we should understand the physics of this.

Like every other molecule and process in the cell, ATP must obey the laws of thermal physics.
We can use our ideal-gas picture to illustrate the activation of ATP quantitatively.
ATP provides free energy by its hydrolysis reaction, which simply means water is necessary for its decomposition:
\begin{equation}
    \atp + \mathrm{H}_2\mathrm{O}
    \rightleftharpoons
    \adp + \phos
    \label{chematp}
\end{equation}
where ADP is adenosine di-phosphate and P$_i$ is the separated inorganic phosphate.
Although this reaction is sometimes shown as uni-directional, proceeding from ATP to ADP only, every chemical reaction is reversible.
For simplicity, we'll omit water and phosphate from our analysis, which won't affect our conclusions; it is straightforward to include them if desired.

The reaction \eqref{chematp} is extremely slow in the absence of a suitable catalyst, \cite{Berg-2002} which biologically is very important.
If ATP hydrolysis happened rapidly in solution, the reaction would quickly reach its equilibrium point and ATP would no longer store free energy --- see below.
In a biological context, both directions of \eqref{chematp} only occur in the presence of a biological catalyst, typically an enzyme, thus facilitating the coupling of hydrolysis to useful work, such as transport, biochemical synthesis, or locomotion. \cite{alberts2007molecular}

The chemical details are buried in the rate constants, which we will call $\ktd$ and $\kdt$, respectively, for the forward and reverse directions of \eqref{chematp}.
To gain some insight, we write down the mass-action equation for ATP, omitting water and phosphate for simplicity:
\begin{equation}
    \frac{ d \, \conc{\atp} }{dt}
    = \kdt \conc{\adp} - \ktd \conc{\atp} \; ,
    \label{atpdt}
\end{equation}
which has an equilibrium point 
\begin{equation}
    \frac{ \conc{\adp} }{ \conc{\atp} } = \frac{ \ktd }{ \kdt } \; .
    \label{atpeq}
\end{equation}
Because of the proximity of the charged phosphate groups in ATP, as shown in Fig.\ \ref{fig:atp}, it is intuitively expected that this equilibrium will greatly favor ADP, which is indeed the case.
In our simplified description \eqref{atpdt}, this means that $\ktd \gg \kdt$.
This great imbalance is due to the chemical details.

The ``activation'' of ATP is not due to the tendency for hydrolysis \emph{per se} but rather because the reaction \eqref{chematp} is kept so far from equilibrium in the cell. \cite{phillips2012physical,chowdhury2013stochastic,zuckerman2010statistical}
That is, ATP does not intrinsically store free energy.
After all, without external input of energy, the reaction \eqref{chematp} will go to equilibrium --- and no free energy will be stored, as in our discussion of ideal gases and transporters.
Instead, the cell continually uses energy from the metabolism of glucose to synthesize ATP, \cite{alberts2007molecular} in turn making the \emph{cellular} concentration ratio much smaller than the equilibrium point \eqref{atpeq}.
\emph{It is in this sense that ATP is activated; it is significantly displaced from equilibrium.}
In equilibrium, by contrast, no free energy is stored regardless of chemical details.

In sum, ATP cannot be understood without physics, but that physics is very basic and accessible.




\begin{acknowledgments}
The author is grateful for support from the National Science Foundation, under grant MCB 1715823.
I very much appreciate helpful discussions with August George, Michael Grabe, Phil Nelson, and John Rosenberg.

\end{acknowledgments}


\appendix
\section{Justifying the ideal gas model}
\label{app:ideal}

Why are we justified in using a ``gas'' formulation in the first place when our particles (molecules or ions) are embedded in aqueous solution?
This is an approximation, of course, but what has been assumed?
First, it is legitimate to focus on only a subset of molecules (omitting water, for example) so long as we correctly account for excluded degrees of freedom via effective interactions, governed formally by the potential of mean force (PMF) \cite{zuckerman2010statistical, mcquarrie2004statistical}.
So our approximation is that the PMF among particles of interest is constant, with zero inter-particle force.
But if some of our particles are (charged) ions, is this reasonable?
One argument is that we are only attempting to learn qualitative features of these transporter systems.
Thus we follow the usual ``spherical cow'' physics strategy.

From a more fundamental physics point of view, the fact is that ``integrating out'' intermediary solution molecules significantly \emph{decreases} effective/PMF interactions in typical cases.
No doubt you are already familiar with the high dielectric constant of water, $\epsilon \approx 80$, but consider carefully what it means.
If charges $q_1$ and $q_2$ separated by a distance $r$ are placed in water, the \emph{effective} Coulomb energy of interaction changes from $q_1 q_2 /r$ to $q_1 q_2 /\epsilon r$, decreasing by almost two orders of magnitude.
In fact, the \emph{direct} Coulomb interaction between the charges does not change at all, but rather the additional interactions between the charges and water reduces the average force.
If water electrostatics and conformational motions were included explicitly, using $\epsilon=1$ would lead to the same observed behavior --- namely, forces weakened by a factor of $\epsilon$ --- among the non-water charges. \cite{zuckerman2010statistical}

The phenomenon of electrostatic \emph{screening} resulting from a mixture of positive and negative mobile ions is even more dramatic. 
Excess ions exponentially damp Coulombic interactions, fundamentally breaking the ``long-ranged'' inverse-distance dependence. \cite{mcquarrie2004statistical, zuckerman2010statistical}
The approximate Debye-H\"{u}ckel potential behaves as $q_1 q_2 e^{-\kappa r} /r$ with a ``screening length'' $1/\kappa \sim 1$ nm in physiological conditions.
This length, in turn, is much less than the typical distance between charged molecules in a cell, suggesting that electrostatics play a significantly smaller role than might be expected, at least in terms of the solution behavior which governs the free energy of interest here.

Qualitatively, the key point is that multi-molecular systems typically self-adjust in a way that reduces mutual interactions among any subset of the system.
So the ideal approximation is much less extreme than it seems at first glance, and indeed it underpins essentially all of the long-established, quantitative field of biochemistry. \cite{Berg-2002, cornish2012fundamentals}

\section{Partition-function derivation of the mixture free energy}
\label{sec:partition}

In our analysis, we assumed the free energies for individual components were simply additive terms in the total free energy.
This is not exactly true.
Here we examine the approximation which was implicitly made and the consequences, which turn out to be insignificant.
Ultimately, the conclusions we have drawn are completely accurate in a qualitative sense and even quantitatively reasonable.
Certainly the additivity assumption for the component free energies is no worse than assuming non-interacting molecules in the first place!

We will consider a two-component system because that is sufficient to understand the issues at play.
Specifically, we'll restrict ourselves to a system with $\natot$ A molecules which can freely exchange among inside and outside compartments --- the same setup and notation as was considered in Sec.\ \ref{sec:two-gases}.

To write the partition function, first recall that the partition function generally is a sum/integral over the Boltzmann factor of all possible configurations of the system. \cite{hill1986introduction,mcquarrie2004statistical,zuckerman2010statistical}
In our case, with two ideal gases in separate compartments, we not only have to sum over the coordinate and momentum degrees of freedom as usual, but also over the discrete states represented by different occupancy numbers $\nain$ and $\naout$ of inside vs.\ outside compartments.
That is, we really have a sum over the full-system partition functions $\ztwo(\nain,\naout)$ for every pair of values $\nain$ and $\naout = \natot-\nain$.
We therefore write
\begin{align}
    \ztot_a(\natot, T, \vin, \vout)
     &= \sum_{\nain=0}^{\natot} \ztwo(\nain, \naout, \vin, \vout, T) \nonumber \\
     &= \sum_{\nain=0}^{\natot} \zidl(\nain, \vin, T) \, \zidl(\natot-\nain, \vout, T)  
     \label{zatotzidl}
\end{align}
where we used the fact that the partition function of two independent systems is simply the product of the individual partition functions, which follows from the factorizability of the Boltzmann factor for independent coordinates. \cite{zuckerman2010statistical}
Substituting from \eqref{zidlone} for $\zidl$, we can write the total partition function exactly with an expression that seems unwieldy at first:
\begin{align}
    \ztot_a(\natot, T, \vin, \vout)
     &= \sum_{\nain=0}^{\natot} 
     \frac{1}{\nain!} \lpar \frac{\vin}{\lambda^3} \rpar^{\nain}
     \frac{1}{(\natot-\nain)!} \lpar \frac{\vout}{\lambda^3} \rpar^{(\natot-\nain)} \; .
     \label{zatot}
\end{align}

Before evaluating this expression, let's pause to understand the underlying physics.
We should think of the partition function \eqref{zatot} as a sum over the (un-normalized) probabilities $w$ for each possible $\nain$ value: $\ztot = \sum_{\nain} w(\nain)$.
In other words, the $w$ values give the relative probabilities of the $\nain$ values and hence define a distribution over $\nain$.
The key point is that statistical mechanics predicts a \emph{distribution} of $\nain$ values, each occurring with the appropriate equilibrium probability.
Necessarily, there will be a single largest probability, but fundamentally the distribution governs the observed behavior.

Returning to the equation, we can dramatically simplify \eqref{zatot} if we multiply and divide by $\natot!$ and observe that the sum is exactly of binomial form, leading to
\begin{align}
    \ztot_a(\natot, T, \vin, \vout)
    &= \frac{1}{\natot!} \lpar \frac{\vin+\vout}{\lambda^3} \rpar^{\natot} \; .
\end{align}
By comparison to \eqref{zidlone}, we see that this is simply the partition function for a \emph{single} ideal gas of $\natot$ particles confined to a volume $\vin+\vout$.
Indeed, each ideal, independent particle in our combined system ultimately can access both $\vin$ and $\vout$, so the result makes sense.

Two observations are important before we address the original question about free energy additivity.
First, note that the final free energy doesn't depend on $\nain$ at all.
This is something you should expect because we have effectively ``integrated out'' --- really, summed over --- $\nain$.
More interesting, the final partition function (and hence, free energy) fully accounts for \emph{all possible} $\nain$ values, which are weighted in by their relative probabilities.
That is, in a full statistical mechanics description, the system isn't limited to a single optimal value, as is the case implicitly based on free-energy minimization.

Returning to \eqref{zatotzidl}, we can now understand the precise approximation which has been made.
First, what does summing free energies imply about the underlying partition functions?
Well, note that if a partition function is \emph{exactly} equal to a product of two other partition functions, e.g., $Z_{ab} = Z_a Z_b$, then the free energy is exactly a sum:
$F_{ab} = -k_BT \ln Z_{ab} = -k_BT \ln Z_a +(- k_BT \ln Z_b)$.
On the other hand, the exact expression \eqref{zatotzidl} for the system we're considering is a sum over partition-function products.
When we write the free energy as a simple sum, we are estimating the partition sum in \eqref{zatotzidl} by the \emph{maximum term}, which is a standard approximation in statistical thermodynamics. \cite{hill1986introduction, mcquarrie2004statistical}.
Specifically, our approximation amounts to 
\begin{align}
    \ztot_a(\natot, T, \vin, \vout)
     &\approx \zidl(\nainstar, \vin, T) \, \zidl(\natot-\nainstar, \vout, T)  \; ,
\end{align}
where $\nainstar$ is the value which maximizes the right-hand side of this expression.
Although approximating \eqref{zatot} by a single term seems unreasonable at first, note that since there are $\natot+1$ terms total and each must be less than the maximum, the error in the \emph{logarithm} of the sum required for the free energy should be of lower order than the dominant term in the thermodynamic limit, $\natot \to \infty$.

Most important of all is to realize that \emph{our key results about transporters} --- which have to do with the type of equilibrium points that exist and the thermodynamic driving which is present away from equilibrium --- \emph{are not affected at all by the details of the maximum-term approximation.}
After all, even if we did not make the approximation, there would still be a minimum free-energy point specifying an optimum $\nainstar$ value;
and further, this optimum would exactly correspond to equal inside and outside concentrations in the special case $\vin=\vout$ based on symmetry arguments.
It's clear the essence of our findings would still hold.

\section{Constrained vs.\ Global Equilibrium}
\label{sec:constrained}
It is useful to compare the condition obtained when A is coupled to B by the transporter, namely \eqref{eqconcfour}, to the previous result \eqref{eqconctwo} when A is the only species and hence uncoupled from B.
The coupling of A to B in the presence of the transporter based on \eqref{chemrxn_sym} shifts the resulting equilibrium for A, and \emph{vice versa,} of course.

The A-B coupling means that we have not necessarily obtained the \emph{global} free energy minimum, but what can be termed a \emph{constrained} minimization.
That is, if we consider $\natot, \nbtot, V, T$ as constants, there are two degrees of freedom ($\nain$ and $\nbin$) but we did not allow all possible pairs of these variables.
Because of our transporter, the pair was constrained to lie on a line specified by \eqref{abtot} rather than being able to explore the entire $(\nain,\nbin)$ plane.
The free energy was minimized among the available values on this line, leading to \eqref{eqconcfour}.

What would we have obtained if both degrees of freedom could vary independently?
In that case, each molecular species would separately equilibrate to the equal-concentration point \eqref{eqconctwo}.
This is the global free energy minimum among all $(\nain,\nbin)$ points, which generally won't be accessible for the transporter-coupled case unless $\no=0$.


\section{More advanced perspective on chemical kinetics via trajectories and the Master Equation}
\label{app:master}

Our discussion in the main text examined chemical kinetics solely in terms of ordinary differential equations (ODEs) like \eqref{dbdt} in the mass-action picture.
ODEs, as you may know, are \emph{deterministic} and lead to a unique solution for given initial conditions.
In our case, that would mean the time-evolution of the concentrations $\conc{\ain}, \conc{\aout}$, etc.\ are defined functions of time, as in Fig.\ \ref{fig:conc_vs_time} (even if we couldn't solve the equations analytically).
The deterministic ODE behavior represents the \emph{average} behavior and we do indeed expect this to be unique based on specified initial conditions.

A more microscopic picture of chemical kinetics accounts for different possible outcomes based on the inherent stochasticity of the system, which is nicely illustrated by returning to the system of Sec.\ \ref{sec:two-gases}.
If particles of an ideal gas of A particles are free to move between ``in'' and ``out'' compartments, as in Fig.\ \ref{fig:ideal_two}, then if we imagine making a ``movie'' of the system at time points separated by a fixed time interval $\dt$, we would generate a sequence of of pairs of particle numbers $(\nain, \naout)$.
For example, starting from an equal distribution of particles we might obtain the sequence 
\begin{equation}
    \Big( \nain(t), \, \naout(t)\Big) = (50, 50), (50, 50), (49, 51), (48, 52), (48, 52), (49, 51), \ldots \;.
\label{naseq}
\end{equation}
If we watched our system a second time, a different sequence likely would occur.
That is, there is a \emph{distribution} of possible trajectories just as for any stochastic system \cite{zuckerman2010statistical}, although formalizing the trajectory picture is well beyond the scope of this article.

The (chemical) ``master equation'' (CME) description is intermediate between the ODE and trajectory pictures, though closer to the latter.
The CME assumes an ensemble picture where multiple independent systems are studied simultaneously, characterized by a time-varying \emph{distribution} of discrete states $p(\nain, \naout; t)$.
The $p$ for each state is simply the fraction of systems in that state, implying the normalization $\sum_{\nain=0}^{\natot} p(\nain,\naout) = 1$ at all $t$.
The CME picture can be understood from trajectories such as \eqref{naseq}.
Imagine generating many one-step trajectories started from the state (50, 50).
We could estimate the distribution of outcomes by counting the resulting states, noting that for a very small time step, only states reachable by translocating a single A particle (or none) occur as in trajectory \eqref{naseq}.
Mathematically, we can encode this in a differential equation which, for the state (50, 50), is given by
\begin{equation}
    \frac{d \, p(50,50)}{dt} = \khatio \, p(51,49) + \khatoi \, p(49,50) - \left( \khatio + \khatoi \right) p(50,50) \; ,
    \label{cmefifty}
\end{equation}
where the circumflex is used to distinguish these single-particle rate constants from the co-transport counterparts $\kio$ and $\koi$ of \eqref{dbdt}.
The different terms in \eqref{cmefifty} account for systems from the ensemble which both arrive at (positive sign) or leave (negative) the state $(50,50)$.
Systems which remain in the same state do not alter the probability.

To obtain a general CME, we let $m$ and $n$ denote state indices -- that is, each state like (50, 50) has a \emph{single} index -- with $\kmn$ and $\knm$ the associated transition rates ($\kmn = k_{m \to n}$), obtaining
\begin{equation}
    \frac{dp_m(t)}{dt} = \sum_{n \neq m} \knm \, p_n(t) - \sum_{n \neq m} \kmn \, p_m(t) \;.
    \label{cme}
\end{equation}
This equation says that the probability of state $m$ increases from incoming transitions and decreases from outgoing transitions, just as in \eqref{cmefifty}.
Note that some rate constants may be strictly zero, for processes which cannot occur in a single step.
The CME governs the time evolution of the distribution over states and allows for outcomes besides the average behavior of a simple kinetics description. \cite{nelson2015physModels}
Note that the the set of rate constants of the CME can be used to generate trajectories consistent with \eqref{cme} via the stochastic simulation  (Gillespie) algorithm. \cite{gillespie2007stochastic,chen2018stochastic}

The CME description unpacks the average behavior of simple chemical kinetics, similar to the way that statistical mechanics is the microscopic theory for thermodynamics.
Note, however, that the discrete states in the CME picture are themselves averages over configurational coordinates treated in statistical mechanics.
Part of the challenge, and beauty, of theory is appreciating the relationship between different levels of averaging.

\section{Questions and exercises for enrichment}
We have seen the very basics of biochemical physics for understanding cellular processes.
Readers may be interested in further issues.

\begin{enumerate}
    \item Apply the analysis above to an antiporter governed by \eqref{chemrxn_anti}.  Write down the free energy and solve for the equilibrium point.  Also write down the governing kinetic equations and show these have the same equilibrium point as the thermodynamic calculation.
    \item Derive the equilibrium effects of varying stoichiometry.  For example, assume two A molecules (ions) are needed to transport a single B molecule.  What relation now holds in place of \eqref{eqconcfour}?
    \item How would the formulation given have to be modified to account for electrostatics?  Assume that A and B are charged molecules and there is a potential difference $\Delta \phi$ between inside and outside.  Further assume charges are sufficiently well screened (see Appendix \ref{app:ideal}) so there are no direct molecule-molecule, or ion-ion, interactions.  What equilibrium relation now holds in place of \eqref{eqconctwo}?
    \item A thermodyanmic explanation was provided for the driving force available from ATP in the cell.  Look up an ATP-driven process in a cell biology book and make a kinetic argument for the driving -- i.e., suggest the sequence of events likely to happen given different  ATP, ADP concentrations and knowledge of the equilibrium point.
    \item Look up the cellular processes that are thought to employ kinetic proofreading and discuss what properties they share and why the proofreading might be important.
    \item Numerous numerical experiments can arise from the material.  A comparison could be made of the CME stochastic formulation (Appendix \ref{app:master}) to the deterministic chemical kinetics prescription; see Ref.\ \onlinecite{chen2018stochastic}.  More microscopically, a particle-based simulation \cite{andrews2016smoldyn} could be performed to test the validity of mass-action assumptions with varying parameters such as density, diffusion and reaction rates.
\end{enumerate}

\clearpage
\section{Figure Captions}

\begin{figure}
    \centering
    \includegraphics[width=6cm]{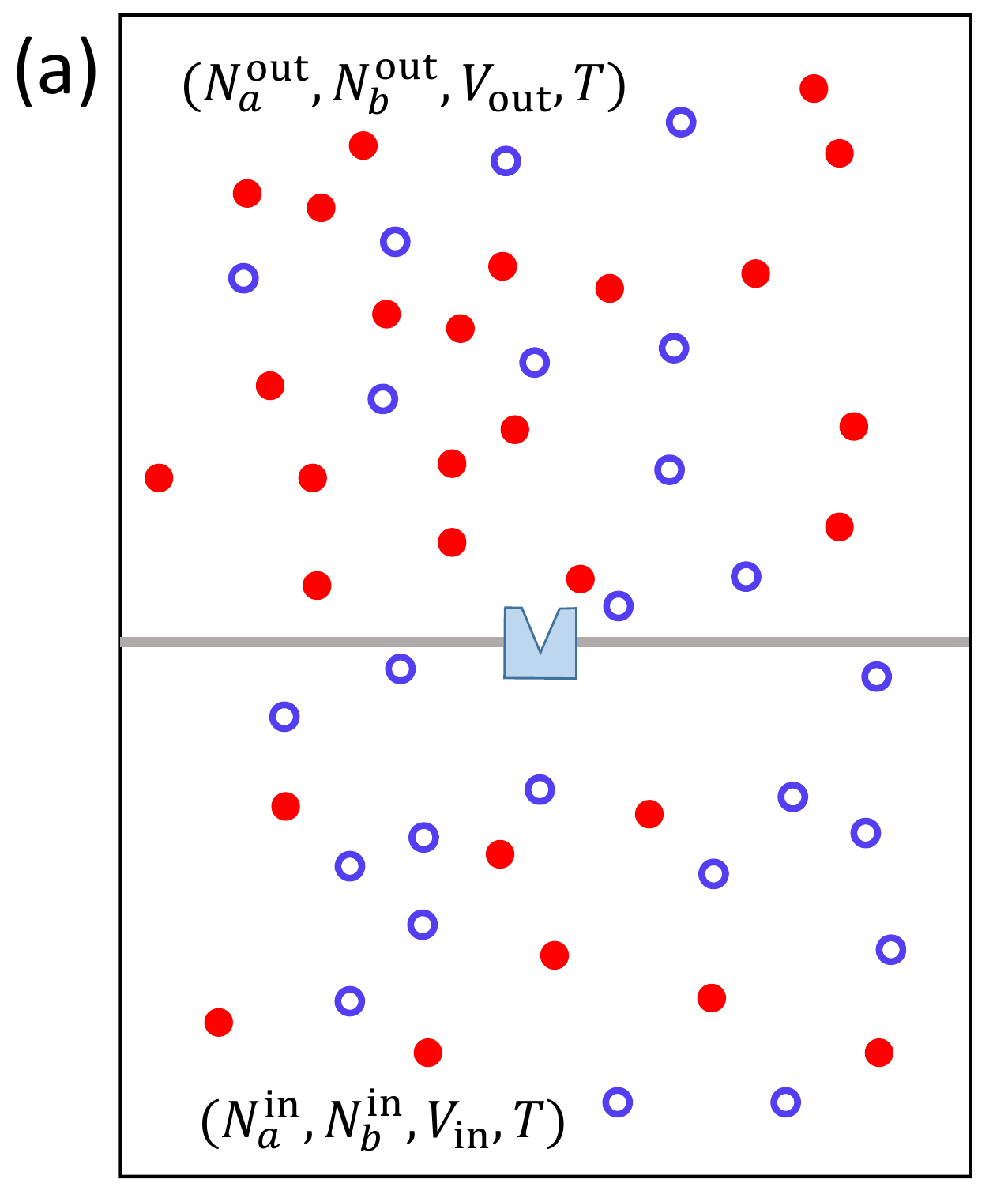}
    \hspace{1cm}
    \includegraphics[width=7cm]{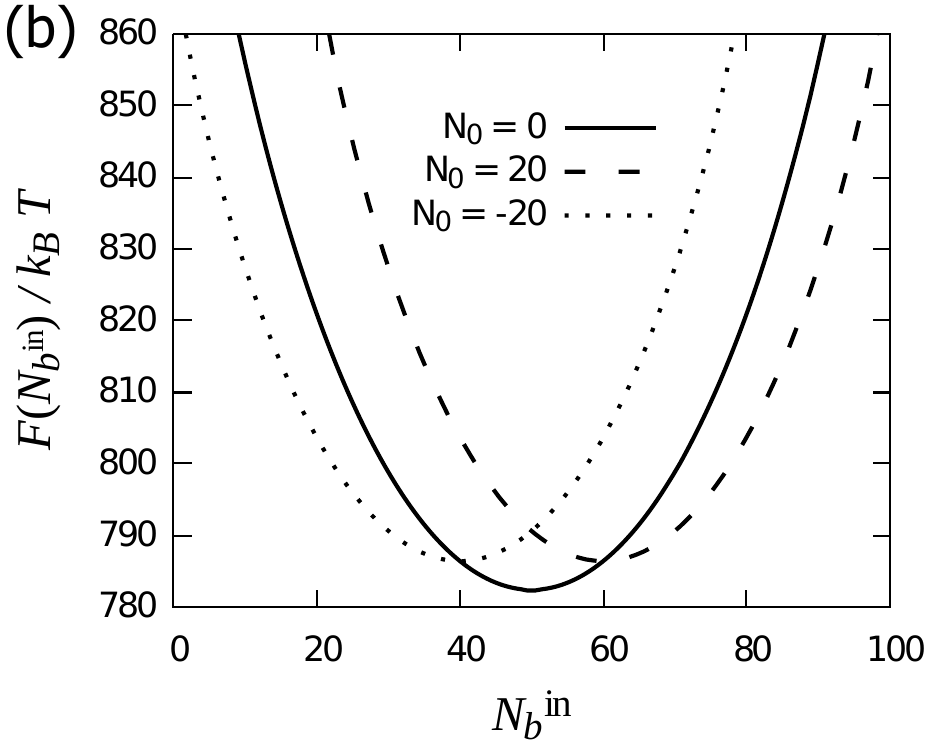}
    \caption{Two ideal gases constrained by a transporter. (a) A mixture of A (red filled circles) and B (blue open circles) particles occupy two compartments separated by an impermeable membrane (gray) with a single embedded transporter (light blue).  The transporter allows free passage of A and B particles, but only in a 1:1 ratio.  (b) The free energy for the system is plotted as a function of the single ``free parameter'' $\nbin$ which self-adjusts to minimized the free energy.  The free energy is shown for different $\no = \nbin - \nain$ values, each of which leads to a different minimum --- i.e., constrained equilibrium.  We have set  $\natot=\nbtot=100$, $\vin=\vout$ and $\vin/\lambda^3 =1$ for convenience.}
    \label{fig:ideal_four}
\end{figure}

\end{document}